\documentclass[twocolumn,showpacs,preprintnumbers,amsmath,amssymb,superscriptaddress,prl]{revtex4-1}
\usepackage{epsfig}
\usepackage{graphicx}
\usepackage{color}
\usepackage{bm}

\begin{document}

\title{Electrical detection of magnetization reversal without auxiliary magnets}

\author{K.~Olejn\'{\i}k}
\affiliation{Institute of Physics ASCR, v.v.i., Cukrovarnick\'a 10, 162 53
Praha 6, Czech Republic} 
\author{V.~Nov\'ak}
\affiliation{Institute of Physics ASCR, v.v.i., Cukrovarnick\'a 10, 162 53
Praha 6, Czech Republic} 
\author{J.~Wunderlich}
\affiliation{Institute of Physics ASCR, v.v.i., Cukrovarnick\'a 10, 162 53 Praha 6, Czech Republic}
\affiliation{Hitachi Cambridge Laboratory, Cambridge CB3 0HE, United Kingdom}
\author{T.~Jungwirth}
\affiliation{Institute of Physics ASCR, v.v.i., Cukrovarnick\'a 10, 162 53
Praha 6, Czech Republic} 
\affiliation{School of Physics and
Astronomy, University of Nottingham, Nottingham NG7 2RD, United Kingdom}

\begin{abstract}
First-generation magnetic random access memories based on anisotropic magnetoresistance required magnetic fields for both writing and reading. Modern all-electrical read/write memories use instead non-relativistic spin-transport connecting the storing magnetic layer with a reference ferromagnet. Recent studies have focused on electrical manipulation of magnetic moments by relativistic spin-torques requiring no reference ferromagnet. Here we report the observation of a counterpart magnetoresistance effect  in such a relativistic system which allows us to electrically detect the sign of the magnetization without an auxiliary magnetic field or ferromagnet. We observe the effect in a geometry in which the magnetization of a uniaxial (Ga,Mn)As epilayer  is set either parallel or antiparallel to a current-induced non-equilibrium spin polarization of carriers. In our structure, this linear-in-current magnetoresistance reaches 0.2\% at current density of $10^6$~A cm$^{-2}$.

\end{abstract}
\maketitle

The first generation of magnetic random access memories (MRAMs) based on the anisotropic magnetoresistance (AMR) relied on magnetic fields for both writing and reading the information in a uniaxial ferromagnet \cite{Daughton1992}. In these AMR-MRAMs, magnetization was reversed by Oersted fields generated by the memory circuitry. The Oersted field was also employed for partially tilting the moments during read-out and by this for breaking the symmetry  between the positive and negative magnetization states. This symmetry breaking made it possible to generate a fraction of a percent AMR signal allowing the detect of the magnetization reversal in a ferromagnetic film. Without the tilt, i.e., for moments flipped strictly by 180$^\circ$, the AMR would vanish. 

In giant magnetoresistance (GMR) or tunneling magnetoresistance (TMR) MRAMs, the auxiliary Oersted field is removed from the read-out scheme \cite{Chappert2007}. Instead, the symmetry breaking is provided by interfacing the storing free ferromagnet with a reference fixed ferromagnet. A spin-dependent transport between the two ferromagnets results in the non-relativistic GMR(TMR) effect with high/low resistance state corresponding to the antiparallel/parallel magnetizations of the free and fixed ferromagnets. In these ferromagnetic bilayer structures, non-relativistic spin transfer torque effects \cite{Ralph2008}, again generated by  the spin-dependent transport between the two ferromagnets, can be also used to remove magnetic fields from writing the free magnet.

Recently, it has been demonstrated that relativistic spin torques can provide means to electrically manipulate ferromagnets without transferring angular momentum between the storing ferromagnet and the auxiliary reference ferromagnet \cite{Miron2011b,Liu2012}. Since the relativistic spin-orbit interaction couples the electron's momentum and spin it can lead to a range of effects when systems are brought out of equilibrium by applied electric fields. Non-equilibrium spin polarization phenomena may occur even in non-magnetic spin-orbit coupled conductors. Prime examples are the inverse spin galvanic effect (ISGE) and the spin Hall effect (SHE) which were experimentally discovered as companion effects, originally in GaAs-based non-magnetic semiconductor structures  \cite{Kato2004b,Kato2004d,Wunderlich2004,Wunderlich2005}. In the ISGE, a non equilibrium spin density  of carriers is generated  in spin-orbit coupled systems which lack inversion symmetry \cite{Silov2004,Kato2004b,Ganichev2004b,Wunderlich2004,Wunderlich2005,Ivchenko2008,Belkov2008}. In the SHE, an electrical current passing through a material with relativistic spin-orbit coupling can generate a transverse pure spin-current polarized perpendicular to the plane defined by the charge and spin-current \cite{Dyakonov1971,Murakami2003,Sinova2004,Kato2004d,Wunderlich2004,Wunderlich2005,Sinova2014}. 

Relativistic spin torques acting in ferromagnets are considered to be related to the ISGE or SHE.  In one picture, a non equilibrium spin density  is generated  via the ISGE and the corresponding effective field induces  the spin torque on the magnetization \cite{Bernevig2005c,Manchon2008,Chernyshov2009,Miron2011b}. Ferromagnetic semiconductor structures, based again on GaAs, provided the initial  experimental evidence of this phenomenon \cite{Chernyshov2009,Fang2011}. In the other picture, a spin current generated  via the SHE propagates from a spin-orbit coupled layer towards the interface with an adjacent ferromagnet where the spin angular momentum of the carriers is transferred to the magnetization \cite{Miron2011b,Liu2012}. 

The remaining challenge for physics, which we focus on here, is to find the counterpart reading scheme for the 180$^\circ$ magnetization reversal in these relativistic structures which does not require a symmetry-breaking magnetic field or a reference ferromagnet. As explained above, AMR is not suitable and the same applies to the recently identified spin Hall magnetoresistance (SMR) \cite{Nakayama2013}. In the SMR,  a spin current generated by the SHE is either absorbed at the interface with a ferromagnet when the SHE polarization is transverse to the magnetization or, in the parallel configuration, it is reflected. The reflected spin current generates an additional voltage via the inverse SHE which renormalizes the resistance of the device. This quadratic-in-SHE phenomenon is, however, independent of the sign of the magnetization. 

In Fig.~1 we demonstrate in our GaAs-based structure that a linear spin Hall magnetoresistance (LSMR) can also occur in these relativistic structures with the maximum and minimum resistance values corresponding to the opposite magnetization directions. The ferromagnet in our structure is represented by an epitaxial 10~nm thick film of Ga$_{0.91}$Mn$_{0.09}$As with Curie temperature $T_{c1}=155$~K. It is grown on top of a 10~nm Ga$_{0.97}$Mn$_{0.03}$As which remains paramagnetic down to $T_{c2}=95$~K, has a similar conductivity that of the top, higher Mn-doped film, and for which we expect a sizable SHE angle \cite{Chen2013,Skinner2015}. This bilayer geometry implies that for the given in-plane current polarity, the SHE generates a fixed in-plane, perpendicular-to-current spin-polarization at the interface with the ferromagnetic (Ga,Mn)As. When the axis of the in-plane magnetization of the ferromagnet is also transverse to the current, as illustrated in Fig.~1(a), then flipping the sign of the magnetization results in the LSMR. The phenomenology can be viewed as analogous to the ferromagnetic  bilayer structure operated in the current-in-plane GMR geometry and with the fixed reference ferromagnet replaced in our structure with the paramagnetic SHE polarizer. Since spin Hall angles, or more generally the charge to spin conversion efficiency,  in strongly spin-orbit coupled systems are of the order of $1-10$\% \cite{Sinova2014},  the LSMR can potentially  have larger amplitudes than the quadratic-in-SHE SMR.

The measurement set-up is shown in Figs.~1(b) and the magnetoresistances detected at 130~K are plotted in Figs.~1(c) and 1(d). The Hall bar of 10~$\mu$m length and 2~$\mu$m width is patterned  by e-beam lithography along the magnetic hard axis ([110] crystal direction) of the ferromagnetic Ga$_{0.91}$Mn$_{0.09}$As film.  At a probe current of amplitude 5~$\mu$A, corresponding to current density of $1.25\times10^4$~Acm$^{-2}$, we observe a negligible change in the longitudinal resistance $R_{xx}$ for the magnetic field swept along the [1$\bar{1}$0] easy-axis, as shown in Figs.~1(c). (All longitudinal resistances showed here and below are scaled by the Hall bar aspect ratio.) On the other hand, at 150 and 300~$\mu$A we observe a magnetoresistance signal which increases with increasing current and is an odd function when reversing magnetization by 180$^\circ$. The hysteretic $R_{xx}$ jumps, whose sign flips when flipping the field sweep direction or the polarity of the probe current, occur at $\pm0.2$~mT, which is the easy-axis coercive field. In Fig.~1(d) we plot  the difference between resistance states for opposite magnetizations, set by sweeping the magnetic field from negative or positive values to the zero field.  The difference increases linearly with the applied probe current and changes sign for the opposite current polarity.

\begin{figure}
\hspace*{0cm}\epsfig{width=1\columnwidth,angle=0,file=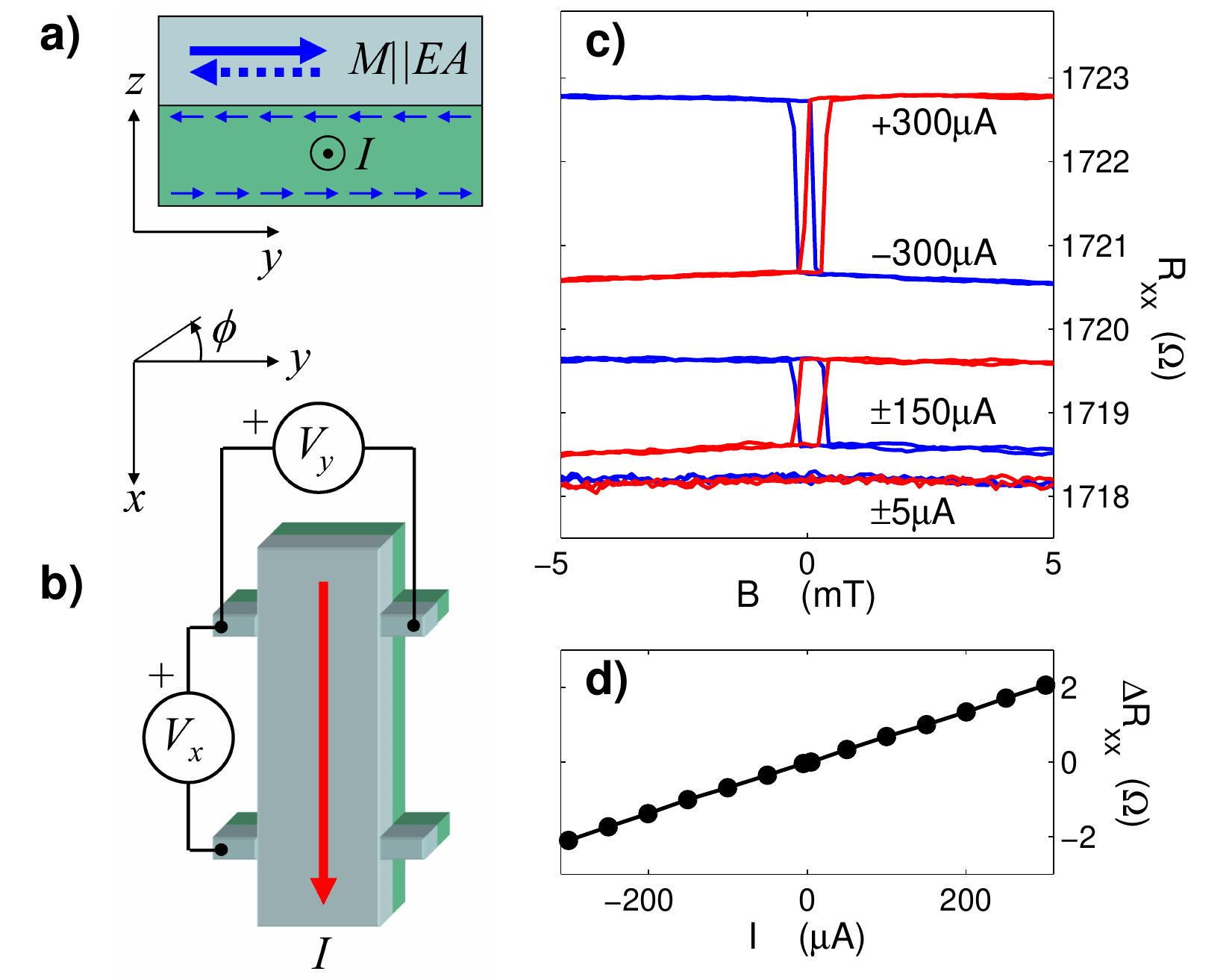}
\caption{(a) Schematic of the LSMR phenomenon. Thin arrows represent the SHE-induced spin polarization; thick arrows represent the easy-axis (EA) magnetization of the ferromagnet. (b) Schematic of the device and measurement geometry. c) Longitudinal resistance measurements at 130~K and different amplitudes and signs of the applied current as a function of the external magnetic field. Steps correspond to the 180$^\circ$ magnetization reversal. (d) Difference between resistance states for opposite magnetizations, set by sweeping the magnetic field from negative or positive values to the zero field, as a function of the applied current.}
\label{fig_device}
\end{figure}

\begin{figure}
\hspace*{0cm}\epsfig{width=1\columnwidth,angle=0,file=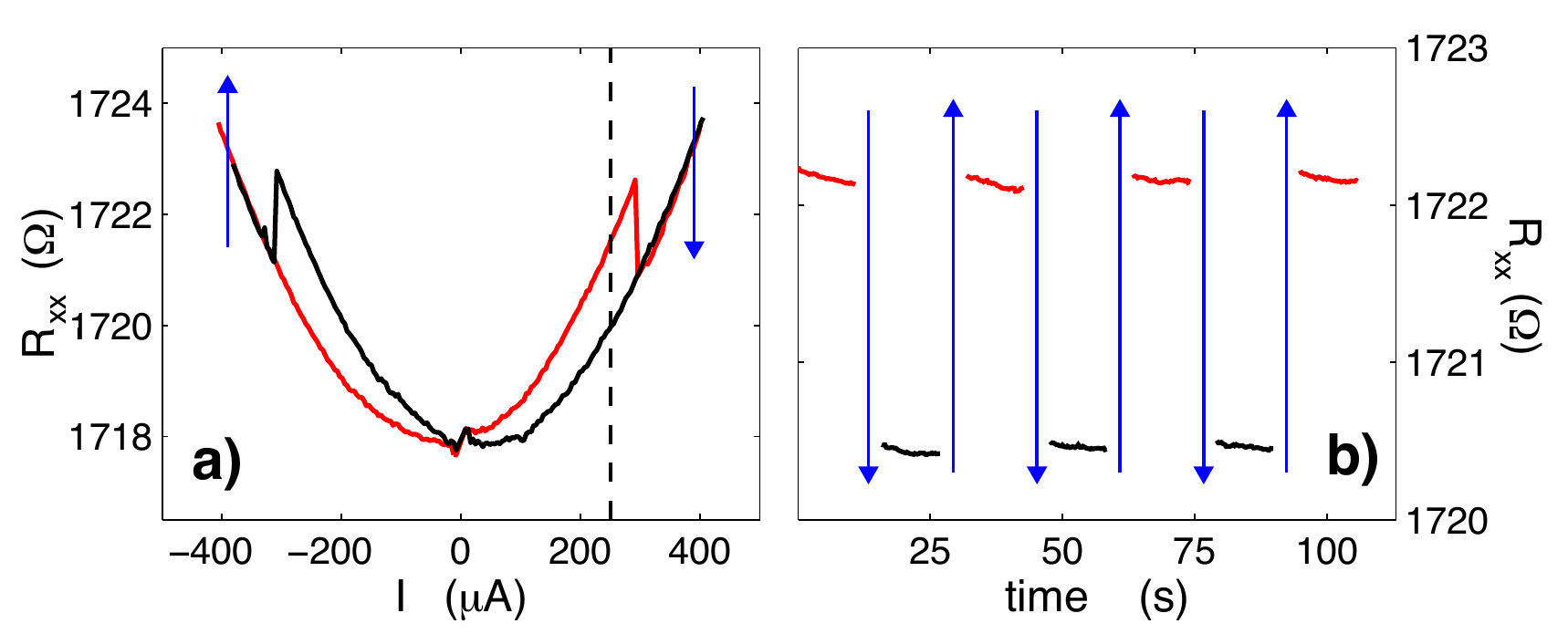}
\caption{ (a) Hysteretic switching measurements in the longitudinal resistance as a function of the applied current at 130~K. Red/black  curves correspond to the up/down current sweeps. (b) All-electrical read/write memory operation of the device. Magnetization switching is induced by setting current pulses of $\mp 400$~$\mu$A, depicted as up/down arrows in both panels. The two states are detected by the $R_{xx}$ measurements at a probe current of amplitude 250~$\mu$A, indicated by the dashed line in panel (a).}
\label{fig_memory}
\end{figure}

The coercive field in our ferromagnetic Ga$_{0.91}$Mn$_{0.09}$As film is sufficiently low that instead of applying the magnetic field externally we can switch the magnetization by the current driven through the Hall bar. This is demonstrated in Fig.~2(a). We estimate that the Oersted field generated in the Hall bar is sufficient to explain the observed switching but the relativistic ISGE or SHE induced fields can have the same symmetry and can contribute to the switching as well. The observed quadratic dependence of $R_{xx}$ on the applied current due to the Joule heating is accompanied by hysteretic jumps  corresponding to the magnetization reversal. In Fig.~2(b) we illustrate the all-electrical read/write memory functionality of our device. The measurements of the memory state are performed in 10~s intervals with the reading current of $+ 250$~$\mu$A. The setting current pulses of length 0.5~s  and amplitude  $\pm 400$~$\mu$A are applied to reverse the magnetization back and forth by  180$^\circ$. We emphasize that the Oersted magnetic field is aligned with the easy-axis of the ferromagnet and, as seen from Figs.~1(c) and 2(a), magnetic fields produce no sizable magnetroresistance outside the switching events and therefore do not facilitated the read-out functionality. We now explore in more detail the mechanism that allows us to electrically detect the opposite magnetization states. 

The magnetoresistance signals observed in Figs.~1 and 2 have the symmetry of the LSMR but, in principle, can also originate from thermal effects. Namely, a vertical temperature gradient can generate  the longitudinal spin Seebeck effect (LSSE) in the paramagnet or the anomalous Nernst effect (ANE) in the ferromagnet \cite{Kikkawa2013b}.  At first glance, one would not expect a thermal effect to depend on the current polarity.  Indeed, the additional thermal voltage due to the LSSE or ANE is polarity independent. However, this additional voltage will enhance or suppress the measured  resistance  depending on the sign of the probe current, precisely mimicking the LSMR symmetry. 

To separate the thermal spin effects we performed angular analysis of the longitudinal and transverse resistances.  The corresponding data are shown in Fig.~3 for in-plane magnetization rotations in a saturating magnetic field of an amplitude 0.5~T and for probe currents, $I\pm300\mu$A. Here $\phi$ is the angle measured from the magnetic easy-axis. The $R_{xx}$ and $R_{xy}$ signals shown in Fig.~3(a) and 3(d) have both a component which is even under the 180$^\circ$ magnetization reversal, as well as a component which is odd under the reversal. The former component is shown in Fig.~3(b) for $R_{xx}$ and in Fig.~3(e) for $R_{xy}$ by plotting the averaged signals, $R_{xx}^{avg}=[R_{xx}(+I)+R_{xx}(-I)]/2$ and $R_{xy}^{avg}=[R_{xy}(+I)+R_{xy}(-I)]/2$. This even component is due to the AMR with its $\sim\cos(2\phi)$ angular dependence for the longitudinal AMR and $\sim\cos(2\phi-45^\circ)$ for the transverse AMR. The odd components are obtained from the differential signals, $R_{xx}^{diff}=[R_{xx}(+I)-R_{xx}(-I)]/2$ and $R_{xy}^{diff}=[R_{xy}(+I)-R_{xy}(-I)]/2$, and are plotted in Figs.~3(c) and 3(f). 

The non-thermal LSMR contributes to the differential component only in the longitudinal resistance (as $\sim\cos(\phi)$). On the other hand, the thermal effects are  present  in the differential signal both for the longitudinal resistance (as $\sim\cos(\phi)$) and for the transverse resistance (as $\sim\sin(\phi)$). Similarly to the non-thermal LSMR, the thermal contribution to the resistance signal  is linear in current since it is obtained from the thermal voltage, proportional to $I^2$, divided by $I$. Since also all $R_{xx}$ data are already rescaled by the Hall bar aspect ratio we can directly  conclude that the amplitude of the thermal signal inferred from Fig.~3(f) is an order of magnitude smaller than the amplitude of the total signal seen in Fig.~3(c). The non-thermal LSMR contribution is then approximately obtained by subtracting $R_{xy}^{diff}$ from $R_{xx}^{diff}$. 

We infer the LSMR contribution with some uncertainty because another linear-in-current term proportional to $\sin(\phi)$ can contribute to the $R_{xy}^{diff}$ signal. It originates from the antidamping-like torque which tends to tilt the magnetization due to a current-induced out-of-plane field and enters $R_{xy}^{diff}$ via the anomalous Hall effect \cite{Avci2014b}. We estimated the strength of the thermal and antidamping-like torque contributions to $R_{xy}^{diff}$  by measuring the $R_{xy}^{diff}$ signals with varying amplitude of the rotating field B (between 0.1T and 2T) and fitting the amplitude of the $R_{xy}^{diff}$ to a function $c_{th}+c_{ad}/(B+B_{aniso})$, where $B_{aniso}$ is the out-of-plane magnetic anisotropy field and $c_{th}$ and $c_{ad}$ are the thermal and antidamping-like torque coefficients, respectively. At 0.5~T, the antidamping-like torque generates about 40\% of the total $R_{xy}^{diff}$ signal. The LSMR signal is, therefore, underestimated when subtracting $R_{xy}^{diff}$ from $R_{xx}^{diff}$. Nevertheless, the error is small since the total $R_{xy}^{diff}$ signal is weak compared to $R_{xx}^{diff}$.

\begin{figure}
\hspace*{0cm}\epsfig{width=1\columnwidth,angle=0,file=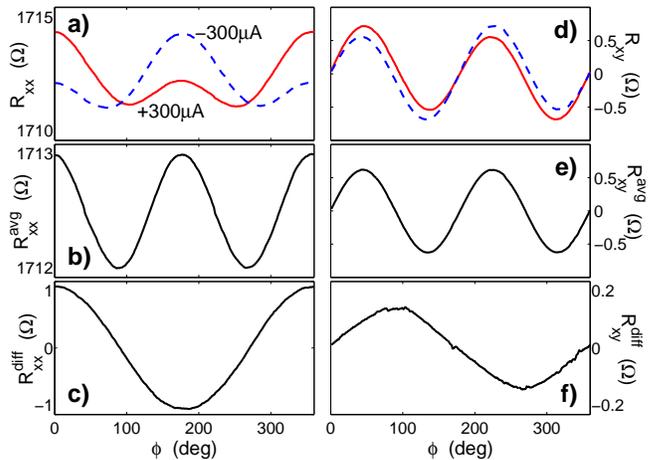}
\caption{In-plane magnetization rotation measurements of the longitudinal resistance signals (a)-(c) and transverse resistance signals (d)-(f) at 130~K, magnetic field of 0.5~T, and applied current of $\pm 300$~$\mu$A.  See text for the definition of the averaged (b,e) and differential (c,f) signals.}
\label{fig_rotation}
\end{figure}

In Fig.~4(a) we show the temperature dependence of the relative amplitudes of the  differential signal $\Delta R_{xx}^{diff}/R_{xx}$,  $\Delta R_{xy}^{diff}/R_{xx}$, and of the non-thermal LSMR contribution $\Delta R_{xx}^{LSMR}/R_{xx}\approx(\Delta R_{xx}^{diff}-\Delta R_{xy}^{diff})/R_{xx}$ obtained from the above rotation experiments in the 0.5~T field. The LSMR component reaches 0.2\% per $10^6$~Acm$^{-2}$ and is significantly larger than the thermal (and antidamping-like torque) component  over a broad interval from $\sim 50$~K to $\sim T_{c1}$ of the top Ga$_{0.91}$Mn$_{0.09}$As film. Remarkably, the LSMR does not disappear below the Curie temperature $T_{c2}$ of the bottom Ga$_{0.97}$Mn$_{0.03}$As epilayer and its sign changes from positive to negative at $\sim 70$~K. The magnetoresistance allowing us to detect the 180$^\circ$ magnetization reversal is present even when the entire  (Ga,Mn)As bilayer becomes ferromagnetic. This is confirmed in Fig.~4(b) where we show measurements in a control device fabricated from a 10~nm thick Ga$_{0.91}$Mn$_{0.09}$As epilayer with no Mn-doping variation and deposited directly on an insulating GaAs substrate. In this sample we also observe the non-thermal 180$^{\circ}$-reversal magnetoresistance signal with the negative sign and with an amplitude comparable to the (Ga,Mn)As bilayer below $T_{c2}$. 

The temperature dependence of the LSMR seen in Fig.~4(a) can be then interpreted as follows: At temperatures below $T_{c1}$  and above $T_{c2}$ the effect has a large magnitude and is due to the SHE in the paramagnetic (Ga,Mn)As layer and the corresponding spin-accumulation at the paramagnet-(Ga,Mn)As/ferromagnet-(Ga,Mn)As interface. The effect persists bellow the Curie temperature of the bottom Ga$_{0.97}$Mn$_{0.03}$As film, which still could be reconciled by recalling that the SHE can in principle also occur in ferromagnets \cite{Miao2013}. However, when the paramagnet/ferromagnet interface is effectively removed below $T_{c2}$ or is not present at any temperature in the control single-layer sample, the remaining  order of magnitude smaller effect is likely of a distinct, bulk origin. Apart from the competition of the SHE and the bulk origins we also note that the resistivity of the higher-T$_c$ ferromagnetic semiconductor material decreases by 30\% between $T_{c1}$ and $T_{c2}$ \cite{Jungwirth2014} which may further contribute to the observed temperature-dependence of the LSMR signal.

Further support for the SHE origin of the large LSMR signal observed at higher temperatures in the paramagnet-(Ga,Mn)As/ferromagnet-(Ga,Mn)As bilayer is provided by comparing data in Figs.~4(a) and 4(c). Here the LSMR effect is measured with the electrical current applied along the [110] and [1$\bar{1}$0] crystal axes, respectively. We observe no significant difference above  $T_{c2}$ between the two measurements, consistent with the SHE picture. 

A competing ISGE mechanism can originate from a broken structural inversion asymmetry at  the bilayer interface or from the inversion asymmetry of the unit cell of the bulk crystal. The former term would generate a Rashba-like ISGE producing a magnetoresistance signal of the same symmetry as the LSMR due to the SHE. However, this structural-asymmetry ISGE term is unlikely to contribute strongly in our epitaxial, iso-structural paramagnet-(Ga,Mn)As/ferromagnet-(Ga,Mn)As bilayer. 

The inversion asymmetry in the bulk crystal of the (Ga,Mn)As epilayers generates ISGE effects of a combined Rashba and Dresselhaus symmetries \cite{Chernyshov2009,Fang2011}. These add up for the current applied along one of the [110] or [1$\bar{1}$0] crystal axes and subtract for the other orthogonal current direction, making the net signal different for the two crystal axes. Consistent with this microscopic picture, the data in Figs.~4(a) and 4(c) develop a clear asymmetry below $T_{c2}$ when the paramagnet/ferromagnet interface is effectively removed. The same asymmetry is also observed between Figs.~4(b) and 4(d) showing the measurements in the control, single-layer (Ga,Mn)As epilayer for currents applied along the [110] and [1$\bar{1}$0] crystal axes.

\begin{figure}
\hspace*{0cm}\epsfig{width=1\columnwidth,angle=0,file=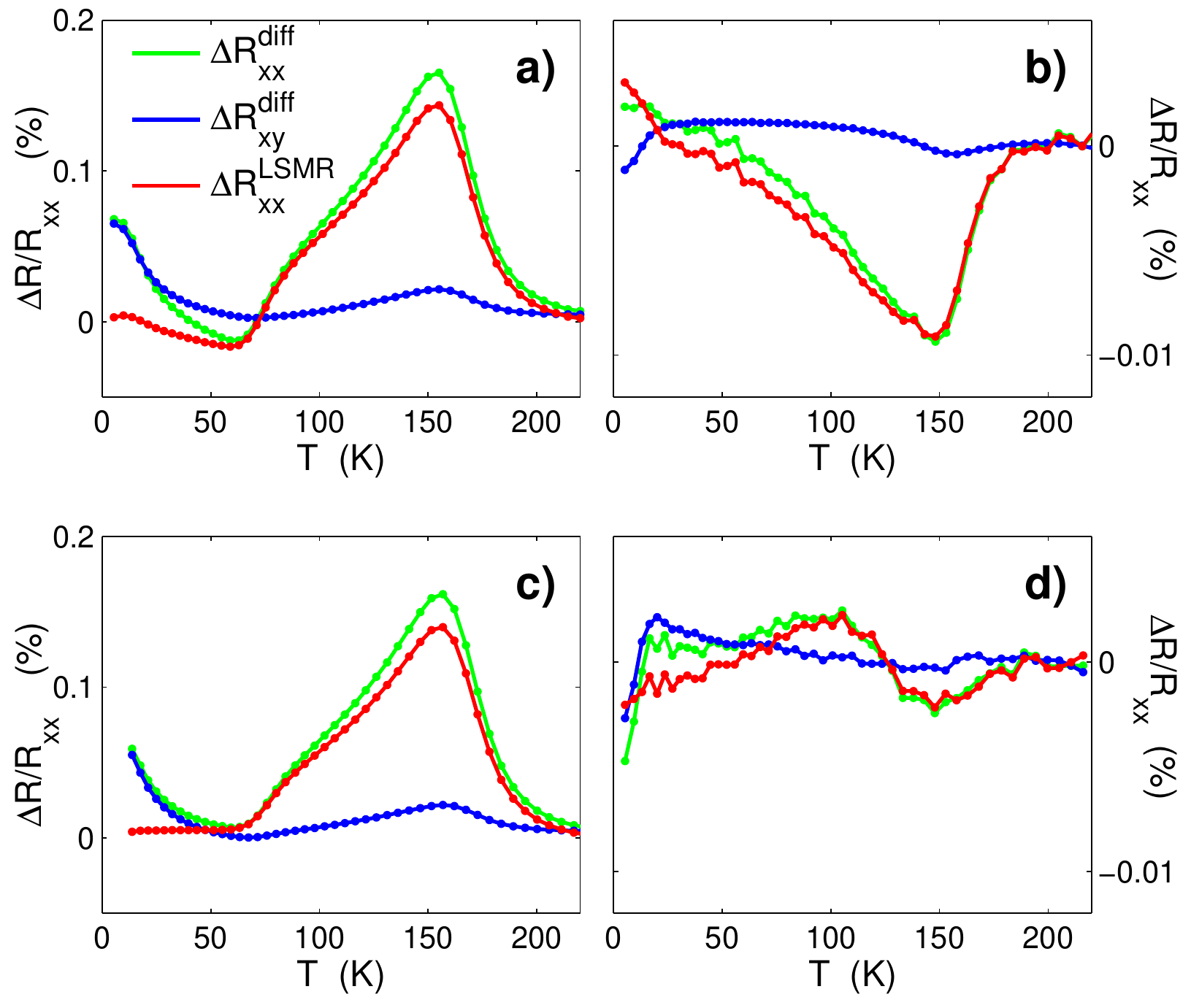}
\caption{(a) Temperature dependence of the relative amplitudes of the longitudinal and transverse differential signals and of the inferred  non-thermal LSMR contribution in the Ga$_{0.91}$Mn$_{0.09}$As/Ga$_{0.97}$Mn$_{0.03}$As bilayer sample for the applied current $\pm300$~$\mu$A (current density $7.5\times10^5$~Acm$^{-2}$) along the [110] crystal axis. (b) Same as (a) in the single-layer Ga$_{0.91}$Mn$_{0.09}$As sample and for the same sign and density of applied current. (c), (d) Same as (a), (b) for current along the [1$\bar{1}0$] crystal axis.}
\label{fig_tempdep}
\end{figure}

To conclude, we have identified a magnetoresistance effect in a paramagnet-(Ga,Mn)As/ferromagnet-(Ga,Mn)As bilayer which allows us to detect a 180$^{\circ}$ reversal in the ferromagnetic film. We attribute the measured signal to the SHE spin-polarization whose direction is fixed by the applied current and which serves as a reference distinguishing the opposite magnetizations in the ferromagnet. Hence, the reference to the GMR geometry with the fixed ferromagnet  being replaced in our structure by the SHE polarizer.  We have also identified a weaker 180$^\circ$-reversal magnetoresistance signal which we attribute to the ISGE in single-layer ferromagnets with bulk inversion asymmetry. This, together with the SHE-induced LSMR in paramagnet/ferromagnet bilayers suggests that the 180$^\circ$-reversal magnetoresistance may occur in a broad class of materials and structures. 

{\em Note added}: Recently we learned about an independent study by Avci {\em et al.} \cite{Avci2015} who reported the observation of a unidirectional spin Hall magnetoresistance in ferromagnet/normal metal bilayers. Several differences can be identified between the two works. In our measurements, the LSMR signal dominates the thermal signal in the $R_{xx}$. Apart from the relative strength of the LSMR compared to the thermal effects, the amplitude of the LSMR  in our semiconductor structure is about two orders of magnitude larger than in the ferromagnet/normal metal bilayer \cite{Avci2015}. We already mentioned in the text that spin Hall angles $\sim 1$\% and correspondingly large spin accumulations  can be expected in the p-type (Ga,Mn)As. Apart from that, our epitaxial paramagnet-(Ga,Mn)As/ferromagnet-(Ga,Mn)As bilayer has an exceptionally high-quality interface. Finally we point out that, albeit highly-doped and degenerate semiconductor, the paramagnet-(Ga,Mn)As has still 2-3 orders of magnitude smaller equilibrium carrier density than common metals. The same current-induced charge to spin conversion efficiency would then generate a significantly larger non-equilibrium spin density relative to the equilibrium charge density and, therefore, a significantly larger net spin-polarization of the carriers which oppose or align with the magnetization, resulting in the stronger LSMR effect.

We acknowledge  support from the EU European Research Council (ERC) Advanced Grant No. 268066, from the Ministry of Education of the Czech Republic Grant No. LM2011026, and from the Grant Agency of the Czech Republic Grant No. 14-37427G.


%

\end{document}